\documentstyle[prl,aps,twocolumn,epsf]{revtex}

\newcommand{\beq}{\begin{equation}}
\newcommand{\eeq}{\end{equation}}
\newcommand{\bea}{\begin{eqnarray}}
\newcommand{\eea}{\end{eqnarray}}

\newcommand{\nn}{\nonumber}
\newcommand{\benn}{\begin{displaymath}}
\newcommand{\eenn}{\end{displaymath}}

\begin{document}

\title{\bf \LARGE Quantum Corrections to Dilute Bose Liquids }

\vspace{0.75cm}

\author{ Paulo F. Bedaque$^1$, Aurel Bulgac$^2$ and Gautam Rupak$^1$ }

\vspace{0.50cm}

\address{$^1$Lawrence-Berkeley Laboratory,
1 Cyclotron Road,  MS 70R0319,  Berkeley, CA 94720-8169}

\address{$^2$ Department of Physics, University of
Washington, Seattle, WA 98195--1560, USA}

\maketitle


\begin{abstract}

It was recently shown [A. Bulgac. Phys. Rev. Lett. {\bf 89}, 050402
(2002)] that an entirely new class of quantum liquids with widely
tunable properties could be manufactured from bosons,
fermions and their mixtures by controlling
their interaction properties by means of a Feshbach
resonance. Since the quantum fluctuations in this regime could in
principle destabilize these objects, we extend the previous mean--field
analysis of these quantum liquids by computing the lowest order
quantum corrections to the ground state energy and the depletion of
the Bose--Einstein condensate and by estimating higher order
corrections as well. We show that the quantum corrections are
relatively small, do not lead to the destabilization of the droplets
and are controlled by the diluteness parameter $\sqrt{n|a|^3} \ll 1$,
even though strictly speaking in this case there is no low density
expansion.

\end{abstract}

\draft

\pacs{PACS numbers:   03.75.Hh, 03.75.Nt, 05.30.Jp }





Among the quantum fluids observed so far only $^3$He and $^4$He and
atomic nuclei are self--bound/liquid systems. It is very likely that
this relatively short list could become extremely long by the addition
of boson droplets (boselets), fermion droplets 
(fermilets) and their mixtures
(ferbolets) \cite{boselets_1}, which owe
their existence to the possibility of tuning two--atom interactions
and to some unusual properties of the three--body system.

It was predicted about ten years ago \cite{feshbach_t} and was
confirmed experimentally about five years ago \cite{feshbach_e} that
by immersing atoms in a magnetic field (and, in principle, in electric
or laser fields as well), one can alter essentially at will the
scattering length $a$ between two atoms. Under such circumstances, it
is possible to create a situation where the scattering length is
negative and significantly larger than the interaction radius.  In the
regime where the magnitude of the scattering length is much larger
than the typical scale of atomic interactions $r_0$, the three--atom
system exhibits some very unusual and interesting phenomena
\cite{efimov_1,efimov_2}. The three--body observables depend on a new
scale $a_3$ which is not determined by any two-body observable.  In the
language of effective field theory, this corresponds to the fact that
a three-body force appears at leading order of the low energy
expansion \cite{stoodges}.  Also, three--body observables are periodic
functions of $\ln (a/a_3)$ and the three--particle scattering
amplitude can be made arbitrarily large by fine--tuning $a$. By
choosing $a$ negative, large ($|a|\gg r_0$) and
with a three--body bound state close to the threshold, a rather unique
situation is created: two low energy atoms experience an effective
attraction, but at the same time three atoms will experience an
effective repulsion. This situation is very well understood
theoretically and it is known as the Efimov effect
\cite{efimov_1,efimov_2}.  A large ensemble of such atoms can condense
in the same way as water droplets condense from vapor.  Since such
condensation can occur for any atom species for which a Feshbach
tunable resonance exist, an entirely new class of (bose, fermi, or
mixtures) quantum liquids \cite{foot_0} can be created.  Their
equilibrium density is determined by the interplay between the
two--body attraction and the three--body repulsion between particles
in the Efimov regime described above. Since the magnitudes of the
two--body attraction and of the three--body repulsion are to a
significant extent under experimental control, the basic properties of
these new quantum liquids are widely tunable \cite{footnote}.

In Ref. \cite{boselets_1}, the basic properties of this new class of
quantum fluids were established in the mean field approximation.  It
is not at all obvious that quantum fluctuations could not destabilize
a boselet. It is generally expected that the mean field approximation is
rather accurate for dilute systems
\cite{pethick,ly_etal,gorkov}. Under normal circumstances (when $a>0$
and $a \approx r_0$) it is sufficient to take into account only the binary
atom--atom collisions and the effects of quantum fluctuations, 
whereas the triple
atom and higher collisions are rather small and controlled by the
diluteness parameter $\sqrt{na^3}\ll1$
\cite{pethick,ly_etal,gorkov}. The regime $a<0$ was considered until
recently as intrinsically unstable towards collapse, unless the number
of bosons is smaller than about 1,500 or so in a trap
\cite{pethick,hulet}.  In the specific Efimov regime, we are interested
in the fact that
the two--body scattering amplitude is large ($a<0,\; |a|\gg r_0$),
but the three--body amplitude is even larger.  Even though the system
is dilute with respect to the two--body collisions ($n|a|^3\ll 1$), it is
not obvious that diluteness with respect to the three--body collisions is
achieved and one might suspect that the mean--field approximation
could be violated. It is imperative to understand the character and
the magnitude of these corrections to the ground state properties in
the case of these new quantum liquids and determine whether quantum
fluctuations can lead to instabilities. One can naturally expect that
such effects would perhaps manifest themselves particularly strongly
in the case of bosons and less so in the case of fermions. For this
reason, we shall focus our attention here on Bose systems only.

We start with the Lagrangian density
\bea
& & {\mathcal  L} + \mu {\mathcal N} =
\psi^\dagger \left (
 i\hbar \frac{\partial}{\partial t}
 +\frac{\hbar^2\bbox{\nabla}^2}{2m} + \mu
\right ) \psi \nn \\
& & -\frac{g_2}{2}\psi^\dagger\psi^\dagger \psi\psi
-\frac{g_3}{6}\psi^\dagger\psi^\dagger\psi^\dagger\psi\psi\psi  ,\label{L}
\eea
where $\psi^\dagger$ and $\psi$ are the creation and annihilation
operators for a boson, $\mu $ is the chemical potential, and the
coupling constants are
\bea
 g_2 &=& \frac{4\pi \hbar^2 a}{m}, \label{g2} \\
 g_3 &=&\frac{12\pi \hbar^2 a^4}{m}
\left [ d_1 +d_2\tan \left ( s_0 \ln \frac{a}{a_3}+
\frac{\pi}{2} \right ) \right ] \nn \\
& =& \frac{6\pi \hbar^2 a^4}{m}\Upsilon  \ .\label{g3}
\eea
Here $a$ is the two--body scattering length and $s_0\approx 1.00624$. $d_1$
and $d_2<0$ are universal constants, whose numerical values have been
determined recently \cite{braaten}.  The values of the coupling
constant are determined by considering the scattering of two and three
particles at zero momentum.  $a_3$ is the value of the two--body
scattering length for which a three--body bound state has exactly zero
energy. Unlike $d_{1,2}$ and $s_0$, the parameter $a_3$ is system
dependent and is also a genuine three--body characteristic.  The rest
of the symbols have their usual meaning.
 We have introduced a new dimensionless quantity $\Upsilon\gg 1$ in
Eq. (\ref{g3}), which will prove very convenient in the power
counting.  We will only use the Lagrangian in Eq.~(\ref{L}) for
particle momenta much smaller than $\hbar/|a|$ so that it is legitimate to
subsume the complicated dynamics on the scale $\hbar/a$, leading to the
Efimov effect into a contact three--body interaction.  We show below
that the typical loop momenta are of order $Q\sim
\hbar/(|a|\Upsilon^{1/2})\ll \hbar/|a| $. More precision can be
systematically attained by including terms in Eq.~(\ref{L}) with more
derivatives or fields, but their effect is strongly suppressed.  As we
mentioned above, the interesting regime is when $g_2<0$ and $g_3>0$
\cite{boselets_1}.

Since we are interested in the condensed state of a system of bosons,
it is natural to split the field $\psi$ into a classical ($c$--number)
condensate $\phi$ and the fluctuations $\psi$: $\psi\rightarrow
\psi+\phi$. ($\phi$ can be chosen real in this case.)  In terms of the
new variables, the Lagrangian density becomes
\bea
& & {\mathcal L} + \mu {\mathcal N}= 
-\frac{1}{2}g_2\phi^4-\frac{1}{6}g_3 \phi^6 +\mu \phi^2+ \nn \\
& &  \phi (\psi+\psi^\dagger)
\left ( \mu - g_2\phi^2 -\frac{g_3}{2}\phi^4\right ) + \nn \\
& & \frac{1}{2}(\psi^\dagger \psi )
\left( \begin{array}{cc}
 i\hbar \frac{\partial}{\partial t} + 
  \frac{\hbar^2\bbox{\nabla}^2}{2m}+\chi
&    - g_2 \phi^2 - g_3\phi^4 \\
     - g_2\phi^2 - g_3\phi^4  
& - i\hbar \frac{\partial}{\partial t} +\frac{\nabla^2}{2m}+\chi \\
      \end{array} \right)
  \left (   
    \begin{array}{c}   
        \psi \\ \psi^\dagger \\
    \end{array} \right) \nonumber \\
& & + \quad {\rm (interactions)}, \label{Lshifted}
\eea
where $\chi=g_2\phi^2+g_3\phi^4$.
The interaction terms include vertices with three $\psi$'s ($\sim g_2
\phi$ and $\sim g_3 \phi^3$), four $\psi$'s ($\sim g_2$ and $\sim g_3
\phi^2$), five $\psi$'s ($\sim g_3 \phi$), and six $\psi$'s ($\sim
g_3$) and normal ordering is implied as well.  At tree level/mean
field, the particle number density is $n=\phi^2$ and the energy density
and chemical potential (determined by imposing the condition
$\langle\psi\rangle=0$) are given by
\bea
& & {\mathcal E}= \frac{g_2}{2}n^2 + \frac{g_3}{6}n^3 ,\\
& & \mu = \frac{d {\mathcal E}}{d n}= g_2 n +\frac{g_3}{2}n^2. \label{chem}
\eea
In writing down the contribution quadratic in $\psi$'s in
Eq. (\ref{Lshifted}) we have implicitly taken into account the
condition (\ref{chem}). At equilibrium/zero pressure 
($P=nd {\mathcal E}/dn - {\mathcal E}=0$)  the values for the
particle number density, energy density, and chemical potential are
easily determined to be
\bea
& & n_0 = \frac{1}{|a|^3} \frac{1}{\Upsilon}, \label{dens_0}\\
& & {\mathcal E}_0= \mu_0 n_0 =
 -\frac{\hbar^2}{ma^2} \frac{\pi}{\Upsilon} n_0 ,\\
& & \mu_0 = \left . \frac{ d {\mathcal E}}{d n}\right |_{n=n_0}
   = \frac{{\mathcal E}_0}{n_0}
   =-\frac{\hbar^2}{ma^2} \frac{\pi}{\Upsilon}. \label{chem_0}
\eea
Notice that only at equilibrium the chemical potential becomes equal
to the energy per particle, see Eq. (\ref{chem_0}). The tree
level/mean field approximation is valid when the system is dilute,
which holds as long as $\Upsilon \gg 1$. At and near equilibrium, the
contribution of the two--body and three--body collisions are
comparable in magnitude. The three--body collision term, which in
dilute systems is typically negligible, is important in this case,
because the three particle bound state is close to the threshold and the
particles are in the Efimov regime.

From the Lagrangian (\ref{Lshifted}) we can determine the normal and
anomalous propagators for the Bogoliubov quasiparticle, which can be 
conveniently represented as a matrix
\beq
D_k=\frac{1}{\omega ^2 -\omega_k^2}
\left (
  \begin{array}{cc}
   \omega + \frac{\hbar^2 \bbox{k}^2}{2m} + \chi & - \chi \\
    -\chi & -\omega +\frac{\hbar^2 \bbox{k}^2}{2m}+ \chi
  \end{array}
\right ) ,
\eeq
where $\omega_k$ is the Bogoliubov
dispersion relation for quasiparticle excitations,
\beq
\omega_k=\sqrt{\frac{\hbar^2\bbox{k}^2}{2m}
\left ( 2\chi+\frac{\hbar^2\bbox{k}^2}{2m}\right ) }
\approx \hbar k \sqrt{\frac{\chi}{m}}. \label{sound_waves}
\eeq
Since $g_2<0$, the sound velocity $s=\sqrt{\chi/m}$ for long 
wavelengths is imaginary for small particle number densities, namely when
\beq
n < n_s = \frac{|g_2|}{g_3}= \frac{2}{3}n_0,  \label{spinodal}
\eeq
and the homogeneous matter at these densities ($n<n_s$) is unstable towards
collapse and condensation into denser droplets. For densities $n_s\leq
n\leq n_0$, the pressure is negative and such a system tends to
increase its density and binding energy. If the density is $n>n_0$, the
internal pressure is positive and the system tends to expand towards
equlibrium and lowers its energy, unless external walls exert inward
pressure. Unlike the phonons in liquid $^4$He, the curvature of the
quasiparticle excitation $\omega_k$ as a function of the wave vector
$k$ is positive. As a result these quasiparticles have a finite
lifetime due to decay into two quasiparticles, corresponding to a
width $\Gamma \sim k^5$ \cite{gorkov,belyaev}, since the velocity of a
quasiparticle with finite wave vector $k$ is larger than the sound
velocity.

The quantum fluctuations can now be evaluated at the one--loop level
with usual techniques. We have chosen to use dimensional
regularization for the evaluation of various diverging integrals
\cite{dimreg}. Old fashioned techniques \cite{pethick,ly_etal,gorkov},
such as diagonalization of the quadratic part of the Lagrangian
(\ref{Lshifted}) with the a Bogoliubov transformation and
subsequent evaluation of the corrections to the ground state density
and energy would lead to identical results. The particle density and
the energy density at one--loop level are given by the following
expressions
\bea
& & n=\phi^2+   \frac{1}{3\pi^2}
      \left ( \frac{m\chi}{\hbar^2}\right )^{3/2}, \label{eq:den1}\\
& & {\mathcal E}=\frac{g_2\phi^4}{2} + \frac{g_3\phi^6}{6} +
      \frac{8\hbar^2}{15\pi^2m}
      \left ( \frac{m\chi}{\hbar^2}\right ) ^{5/2}
    + \mu (n-\phi^2). \label{eq:en1}
\eea
The first correction term to the energy density (and similarly to the
number density as well) is formally identical to the Lee and Yang
term \cite{pethick,ly_etal,gorkov}, which was computed, however, for a
Bose gas with repulsive two--body interaction.  The equilibrium values
for the particle number density (both condensed and noncondensed),
the energy density, and energy per particle, after including 
first--order quantum corrections become
\bea
& &  n =  \frac{1}{|a|^3} \frac{1}{\Upsilon}
\left ( 1 -\frac{72}{5}\alpha \right ),
\quad {\mathrm{where}} \quad
\alpha=
  \frac{2^{3/2}}{3\pi^{1/2} \Upsilon ^{1/2} }  , \label{density}\\
& &  {\mathcal E}= {\mathcal E} _0 \left ( 1- \frac{88}{5}\alpha \right),
\quad
\frac{{\mathcal E}}{n} = \mu_0
\left ( 1 -\frac{16}{5}\alpha \right ) .
\eea
Surprisingly, Eqs. (\ref{eq:den1}) and (\ref{eq:en1}) determine
correctly both the linear and the quadratic corrections in $\alpha$ for
energy per particle ${\mathcal E}/n$.  Therefore, the first order
quantum corrections are controlled by the same diluteness parameter
$\sqrt{n|a|^3}=\Upsilon^{-1/2}\ll 1$ as in the case of a Bose gas with
repulsive interaction. In spite of this formal apparent similarity,
the meaning of the present result for the condensed Bose liquid when
$a<0$ is qualitatively different. There is strictly no low--density
expansion in the present case, since for densities $n<n_s=2n_0/3$, the
system is unstable towards long--wavelength density fluctuations as
the sound velocity is imaginary, see discussion of Eqs.
(\ref{sound_waves}) and (\ref{spinodal}).

A general argument can be given for the suppression of higher loop
corrections.  Consider a generic diagram contributing to the energy
density (that is, without external legs) containing I propagators, L
loops and $n_i$ vertices with $i$ legs. It can be estimated by
\bea
& & {\rm (Energy\  Diagram)} \sim \left( \frac{m}{Q^2} \right)^I
     \left( \sqrt{\frac{\chi}{m}}Q Q^3 \right)^L \nn \\
& & \times
     \left ( \frac{\chi}{n^{1/2}} \right ) ^{n_3}
     \left ( \frac{\chi}{n         } \right ) ^{n_4}
     \left ( \frac{\chi}{n^{3/2}} \right ) ^{n_5}
     \left ( \frac{\chi}{n^2     } \right ) ^{n_6} \nn \\
&  & \sim  m^{3L/2}
      \chi^{-I + 5L/2 + n_3 + n_3 + n_5 + n_6} \nn \\
& & \times     n^{ - n_3/2 - n_4 - 3n_5/2 - 2n_6},
\eea
where $Q\sim \sqrt{m\chi}\sim 1/(|a|\Upsilon^{1/2})$ is the typical
loop momentum \cite{foot_1}. When using dimensional regularization
no powers of the cutoff are present and the only momentum scale
in these diagrams is set by $\sqrt{m \chi}$. Combining the above
estimate with the relations
\bea
2 I &=& 3 n_3 + 4 n_4 + 5 n_5 + 6 n_6,   \\
I &=& n_3+n_4+n_5+n_6+L-1,  \ \ \ \ \ \ \ \
\eea
we find
\beq
{\rm (Energy\  Diagram)}\sim m^{\frac{3L}{2}}\chi^{\frac{3L}{2}+1} n^{1-L}
 \sim {\mathcal E}_0 \Upsilon ^{-L/2}.  \label{L_loops}
\eeq
Consequently, up to logarithmic corrections, all higher--order loop
corrections are suppressed in the dilute limit by powers of
$\sqrt{n|a|^3}=\Upsilon^{-1/2}\ll 1$. As in the case of dilute Bose
gases with repulsive interactions \cite{pethick,ly_etal,gorkov,BEC,mott} this
parameter is proportional to the ground--state fraction of non--BEC
particles. In this respect, this new class of quantum liquids, in
particular the boselets, is qualitatively different from liquid
$^4$He, where the fraction of non--BEC particles is close to 90\%.

For large $\Upsilon \gg 1$, the self--bound system is dilute.  The states
with positive pressure, however, can be dense.  These states could
still be dilute with respect to two--body collisions, however of such
a density that $|g_2|\ll g_3 n$. In this regime, the loop expansion
breaks down and a further resummation is required.  The energy per
particle of the system is large and the system can exist only if
external pressure is applied. It is very likely that the system
undergoes a transition to a coherent mixture of atoms and trimers,
similar to the situation put in evidence in gaseous BEC, where the
coexistence of an atomic BEC with a molecular BEC was predicted
\cite{amBEC_t} and experimentally observed \cite{amBEC_e}.  The
formation of trimers will have two effects, lowering the particle
density and lowering the energy.  The spatial size of a trimer when
$a<0$ is smaller than $|a|$, even if the trimer binding energy is
vanishing.  The wave function of an Efimov state of zero energy is
normalizable and the wave function is concentrated predominantly at
interparticle distances $|\bbox{r}_i-\bbox{r}_j|<|a|$, where
$i,j=1,2,3$ are particle labels in a trimer \cite{foot_2}.  Since the
average separation between atoms is presumed to be larger than the
scattering length, when all atoms convert into trimers, the trimer
number density decreases roughly by a factor of 3 when compared to
the atomic number density \cite{boselets_1}. The trimer phase is most
likely stable, provided the trimer--trimer scattering length is
positive, which is most probable \cite{pethick,gribakin}. When a
trimer is formed, a significant amount of energy is released as
well. These two factors, the significant drop in density and energy,
should be explicitly taken into account when evaluating the energy
density at higher particle densities, where triple collisions might
dominate.  The self--bound state we have considered is only
meta--stable. It can decay either by a four particle process, leading
to the formation of a trimer and one energetic atom, or through a
three--body process with the formation of a deeper dimer and one
energetic atom (the formation of a shallow dimer is excluded since
$a<0$). The rate for the first process is proportional to $\hbar n^4
a^7/m $ and likely very small. The formation of a deep dimer was
considered in Ref. \cite{deep_bs} and it was found to be of the order
of $\hbar n^3 a^4/m$. This decay rate is, however, enhanced when a
three--body state is close to the threshold, where the correct rate can be
determined only through a non--perturbative calculation, 
but this is not yet
attempted. This recombination rate is related to the widths of the
Efimov states due to the existence of deep two--body bound states,
which follow the same exponential behavior as the energies of the
Efimov states \cite{nielsen}.

We thank G.F. Bertsch and D.B. Kaplan for discussions.

{\it Note added.} Boselets made of spin
polarized tritium atoms are the likely candidates, since the regime we study
here, when $g_2<0$ and $g_3>0$, could apparently be easily reached
according to Ref. \cite{tritium}. In this particular
case, three--body recombination processes are absent.

{\it Note added in proof.} Recently Braaten and Hammer \cite{eh} have performed
a nonperturbative calculation of three--body recommbination rates in
dilute gases.



\begin{thebibliography}{99}
\bibitem{foot_0} The condensation in this case has a double meaning:
(1) there is a real condensation similar to the phase transition
between a gas and a liquid which occurs in real space, and (2) there is
Bose--Einstein condensation at temperatures close to absolute zero in
momentum space.
\bibitem{foot_1} Notice that the energies scale such as $\sqrt{\chi/m}Q$,
in accord with the Bogoliubov dispersion relation in the small
momentum region. The cutoff may appear inside logs, so we will not be
able to determine the powers of $\ln (na^3)=-\ln \Upsilon $ without a
more detailed argument.
\bibitem{foot_2} The wave function of a trimer with zero energy decays
at large distances ($R\gg |a|$) as $1/R^4$, where $R$ is the
hyperspherical radius. This behavior can be easily inferred from the
behavior of the zero--energy three--particle propagator and by writing
the Schr\"odinger equation as a Lippmann--Schwinger equation or, even
better, as a Fadeev equation for the wave function.
\end{thebibliography}
\end{document}